\documentclass[aps,prd,twocolumn,showpacs,showkeys,amsmath,amssymb]{revtex4}
\usepackage{epsfig} 
\usepackage{makecell}
\usepackage{array}
\usepackage{subfigure}
\usepackage{bm}
\usepackage{longtable}
\usepackage{tabularx}
\usepackage{booktabs}
\usepackage{appendix}
\usepackage{caption2}
\usepackage{amsfonts}
\usepackage{graphicx}
\graphicspath{{figs/}}
\usepackage{dcolumn}
\usepackage{siunitx}
\usepackage{bm}
\usepackage{float}
\usepackage{txfonts}
\usepackage{overpic}
\usepackage{feynmf}   
\usepackage{epstopdf}   
\usepackage{slashed}  
\usepackage{multirow}

\usepackage[colorlinks, citecolor=blue,anchorcolor=red,menucolor=red, linkcolor=red,filecolor=red,runcolor=red,urlcolor=blue,frenchlinks=red]{hyperref}

\usepackage{ulem}
\usepackage{color}

\allowdisplaybreaks[4]

\makeatletter
\newcommand{\figcaption}{\def\@captype{figure}\caption}
\newcommand{\tabcaption}{\def\@captype{table}\caption}

\newcommand{\Rmnum}[1]{\expandafter\@slowromancap\romannumeral #1@}

\def\hlinewd#1{%
  \noalign{\ifnum0=`}\fi\hrule \@height #1 \futurelet
   \reserved@a\@xhline}
\makeatother

\newcommand\dqq{\Big\langle \bar q q \Big\rangle}
\newcommand\dss{\Big\langle \bar s s \Big\rangle}
\newcommand\dqGq{\Big\langle g_s \bar q \sigma G q \Big\rangle}
\newcommand\dsGs{\Big\langle g_s \bar s \sigma G s \Big\rangle}
\newcommand\dGG{\Big\langle \alpha_{s} GG \Big\rangle}
\newcommand\dGGG{\Big\langle g_{s}^{3} G^{3} \Big\rangle}

\begin{document}
\title{Predictions of the hybrid mesons with exotic quantum numbers $J^{PC}=2^{+-}$}

\author{Qi-Nan Wang$^1$}
\author{Ding-Kun Lian$^1$}

\author{Wei Chen$^1$}
\email{chenwei29@mail.sysu.edu.cn}
\affiliation{$^1$School of Physics, Sun Yat-Sen University, Guangzhou 510275, China}

\begin{abstract}
We study the non-strange and strangeonium light hybrid mesons with $J^{PC}=2^{+-}$ by using the method of QCD sum rules. The local hybrid interpolating currents with three Lorentz indices are constructed to couple to such exotic quantum numbers. We calculate the correlation functions up to dimension eight condensates at the leading order of $\alpha_{s}$. In our results, the masses of the non-strange $b_2$ and $h_2$ hybrids are about $2.65~\mathrm{GeV}$, while that of the strangeonium $h_2^\prime$ hybrid is about $2.72~\mathrm{GeV}$. Such exotic $2^{+-}$ hybrids can be generated through both the two-gluon and three-gluon emission processes in the radiative decays of $\chi_{cJ}$. Moreover, these hybrid mesons may be detectable due to their peculiar decay behaviors and small decay widths. Using the high-statistics data samples of $\psi(3686)$ in BESIII and BelleII, it is possible to hunt for such hybrid states through the partial wave analyses in the $b_2\to\omega/a_1/h_1/a_2\pi\to4\pi$, $h_2\to\rho\pi\to3\pi$ and $h_2\to b_1\pi\to5\pi$ processes.
\end{abstract}


\maketitle

\section{Introduction}
As the most successful theory of strong interaction, the quantum chromodynamics (QCD) has predicted the existence of exotic hadron states beyond the conventional quark model, in which hadrons are $q\bar q$ mesons and $qqq$ baryons. Among the exotic hadrons, a hybrid meson ($\bar qgq$) is composed of a pair of quark-antiquark and a valence gluon. Due to the existence of gluon degrees of freedom, hybrid mesons can produce quantum numbers forbidden by the quark model, such as $J^{PC}=0^{\pm-}, 1^{-+}, 2^{+-}$.
In the past several decades, the states with these exotic quantum numbers have received extensive research interest in both theoretical and experimental aspects~\cite{Chen:2016qju,Esposito:2016noz,Guo:2017jvc,Liu:2019zoy,Brambilla:2019esw,Chen:2022asf,Meng:2022ozq}.

There is no solid signal for the existence of hybrid mesons to date. The most fascinating candidates of light hybrid mesons are three exotic quantum number states $\pi_{1}(1400)$~\cite{IHEP-Brussels-LosAlamos-AnnecyLAPP:1988iqi}, $\pi_{1}(1600)$~\cite{E852:2001ikk} and $\pi_{1}(2015)$~\cite{E852:2004gpn} with $I^GJ^{PC}=1^-1^{-+}$, among which the $\pi_{1}(1400)$ and $\pi_{1}(1600)$ were considered to be the same state by the coupled-channel amplitude analysis~\cite{JPAC:2018zyd} according to the COMPASS data of the $\eta^{(\prime)}\pi$ system~\cite{COMPASS:2014vkj}. 
Recently, the BESIII Collaboration reported an isoscalar state $\eta_{1}(1855)$ with $I^GJ^{PC}=0^+1^{-+}$ in the decay process $J/\psi \to \gamma \eta_{1}(1855)\to\gamma\eta\eta^{\prime}$, with the mass and decay width $m=1855 \pm 9_{-1}^{+6}~\mathrm{MeV}$ and $\Gamma=188\pm 18_{-8}^{+3}~\mathrm{MeV}$ respectively~\cite{BESIII:2022riz,BESIII:2022iwi}. 
Since its observation, the $\eta_{1}(1855)$ has been considered as a candidate for hybrid meson\cite{Chen:2022qpd,Chen:2023ukh,Chen:2022isv,Qiu:2022ktc,Shastry:2022upd,Shastry:2023ths,Shastry:2022mhk}, although other interpretations can not be excluded\cite{Huang:2022tpq,Yan:2023vbh,Yang:2022rck}. Furthermore, the $\eta_{1}(1855)$ was assigned as the isoscalar partner of the $\pi_{1}(1600)$ state in a $J^{PC}=1^{-+}$ hybrid nonet~\cite{Chen:2023ukh,Qiu:2022ktc,Shastry:2022upd,Shastry:2022mhk}. 

In the MIT bag model~\cite{Barnes:1982tx,Chanowitz:1982qj}, the hybrid mesons with $J^{PC}=1^{--}, 0^{-+}, 1^{-+}$ and $2^{-+}$ were established to be the lightest supermultiplet states, in which only $1^{-+}$ is the exotic quantum number. The exotic quantum numbers $J^{PC}=0^{+-}$ and $2^{+-}$ were assigned to the higher supermultiplet, while the $0^{--}$ hybrid was much heavier with different gluon excitation~\cite{HadronSpectrum:2012gic,Chen:2013zia,Meyer:2015eta}. Such hybrid supermultiplet structures were supported by the Coulomb gauge QCD~\cite{Guo:2008yz} and LQCD~\cite{Dudek2011,Dudek:2009qf,Dudek:2010wm,Dudek:2011tt} calculations. Besides, the hybrid mesons have been extensively studied by various methods, such as the LQCD~\cite{Lacock1997,Lacock1996,HadronSpectrum:2012gic,Ma:2020bex,Dudek2011,Dudek:2009kk,Dudek2013,Dudek:2009qf,Dudek:2010wm,Dudek:2011tt}, the flux tube model~\cite{Isgur:1984bm,Isgur:1985vy,Burns2006}, Bethe-Salpeter equation~\cite{Burden2002,Burden:1996nh} and QCD sum rules~\cite{Balitsky:1982ps,Latorre:1984kc,Govaerts:1983ka,Govaerts:1984bk,Govaerts:1984hc,Govaerts:1986pp,Balitsky:1986hf,Barsbay:2022gtu,Chen:2021smz,Chen:2022qpd,Chen:2013eha,Ho:2018cat,Ho:2019org,Palameta:2018yce}. 

Comparing to the $1^{-+}$ channel, the hybrid states with exotic $J^{PC}=0^{+-}$ and $2^{+-}$ have received much less theoretical interest, partly because of the absence of experimental signal. However, it is especially interesting to note that the exotic $2^{+-}$ hybrids were predicted to be surprisingly narrow with the decay widths less than 10 MeV in the PSS model~\cite{Page:1998gz}. If this is true, such narrow states would be detectable in the definite hybrid decay modes. 

The hybrid mass is also an important parameter to determine its decay properties. In LQCD~\cite{Dudek2011,Dudek2013}, the masses of the light $2^{+-}$ hybrids were predicted to be about $2.4-2.8 ~\mathrm{GeV}$ with a heavier pion mass than the physical one. Considering the effect of the light quark  masses, the hybrid masses shall be slightly lighter when the pion mass tends to the physical point~\cite{Dudek2011,Meyer:2015eta}.  

In past several decades, QCD sum rule has been extensively applied to study the hadron structures and properties, such as hadron masses, coupling constants, decay widths, magnetic moments and so on~\cite{Shifman:1978bx,Reinders:1984sr,Colangelo:2000dp}. Based on QCD itself, the QCD sum rule formalism is to approach the bound state problem from the asymptotic freedom side at short distances to long distances where confinement emerges and hadrons are formed. It has been proven to be very successful for studying not only the conventional hadrons but also the exotic ones~\cite{Nielsen:2009uh,Narison:2022paf,Chen:2022asf}. The applications of QCD sum rules to hybrid mesons are almost as early as for the conventional baryons and mesons~\cite{Balitsky:1982ps,Latorre:1984kc,Govaerts:1983ka,Govaerts:1984bk,Govaerts:1984hc,Govaerts:1986pp,Balitsky:1986hf}. Nevertheless, there are still some methodological drawbacks of traditional QCD sum rule studies, for example, the absent of high dimension condensates and high order $\alpha_s$ corrections in OPE series, the small pole contribution problem for the multiquark systems, etc. These drawbacks will weaken the reliability and accuracy of the QCD sum rule predictions, which shall be always kept in mind and considered in analyses. There are some efforts recently to discuss these problems~\cite{Pimikov:2022brd,Albuquerque:2020hio,Albuquerque:2021tqd,Wu:2022qwd}.
 
In this work, we investigate the light hybrid mesons with exotic quantum numbers $J^{PC}=2^{+-}$ within the method of QCD sum rule.We shall construct the local hybrid interpolating currents without the covariant derivative operators. The masses of the $2^{+-}$ non-strange $\bar{q}gq$ and strangeonium $\bar{s}gs$ hybrids can be investigated by using these three Lorentz indices currents. Hereafter, we adopt the notation for hybrid mesons following Ref.~\cite{Meyer:2010ku} and the Particle Data Group (PDG)~\cite{ParticleDataGroup:2022pth}, so that the isovector hybrid with $J^{PC}=2^{+-}$ is $b_2$, the isoscalar non-strange one is $h_2$ while the strangeonium one is $h_2^\prime$.
Finally, according to our predicted masses, some experimental suggestions are made. We propose to search for these light hybrid states from the decays of $\psi(3686)$ in future experiments such as BESIII and BelleII.

\section{Currents and projectors}
A hybrid current is usually built out of a $\bar qq$ pair and an excited gluon field. 
As in Ref.~\cite{Govaerts:1986pp}, the hybrid currents with two Lorentz indices can couple to the quantum numbers $J^{PC}=2^{-+}$ and $2^{++}$.
For $J^{PC}=2^{+-}$ and $2^{--}$, we construct the interpolating hybrid currents with three Lorentz indices. 
Using the gluon field strength $G_{\alpha\beta}(x)$ and Dirac matrix, the possible hybrid currents with three Lorentz indices are 
\begin{equation}
 \begin{aligned}
    J_{\alpha \beta \gamma }^{(1)}&=\bar{q} g_s \gamma_\alpha \gamma_5 G_{\beta \gamma} q, ~~~~J^{PC}=(0\, ,1\, ,2)^{\pm-}\, ,\\
        J_{\alpha \beta \gamma }^{(2)}&=\bar{q} g_s \gamma_\alpha \gamma_5 \tilde{G}_{\beta \gamma} q, ~~~~J^{PC}=(0\, ,1\, ,2)^{\pm-}\, ,\\
    J_{\alpha \beta \gamma }^{(3)}&=\bar{q} g_s \gamma_\alpha  G_{\beta \gamma} q, ~~~~J^{PC}=(0\, ,1\, ,2)^{\pm+}\, ,\\
        J_{\alpha \beta \gamma }^{(4)}&=\bar{q} g_s \gamma_\alpha  \tilde{G}_{\beta \gamma} q, ~~~~J^{PC}=(0\, ,1\, ,2)^{\pm+}\, ,\\
      \end{aligned}
  \label{Eq:current}
  \end{equation}
where $q$ is a quark field ($u, d$ or $s$), $g_s$ is the strong coupling constant, and $\tilde{G}_{\alpha\beta}=\frac{1}{2}\varepsilon_{\alpha\beta\mu\nu}G^{\mu\nu}$ is the dual gluon field strength. Thus the interpolating currents $J_{\alpha \beta \gamma }^{(2)}$ and $J_{\alpha \beta \gamma }^{(4)}$ have the opposite parity with $J_{\alpha \beta \gamma }^{(1)}$ and $J_{\alpha \beta \gamma }^{(3)}$, respectively. 

In general, a current with three Lorentz indices can couple to various hadron states via the following relations
\begin{equation}
  \begin{aligned}
\left\langle 0\left|J_{\alpha \beta \gamma}\right| 0^{(-P) C}(p)\right\rangle&=Z_1^0 p_\alpha g_{\beta \gamma}+Z_2^0 p_\beta g_{\alpha \gamma}+Z_3^0 p_\gamma g_{\alpha \beta}
\\&+Z_4^0 p_\alpha p_\beta p_\gamma  \,,\\
\left\langle 0\left|J_{\alpha \beta \gamma}\right| 0^{PC}(p)\right\rangle&=Z_5^0 \varepsilon_{\alpha \beta \gamma \tau} p^\tau  \,,\\
\left\langle 0\left|J_{\alpha \beta \gamma}\right| 1^{PC}(p)\right\rangle&=Z_1^1 \epsilon_\alpha g_{\beta \gamma}+Z_2^1 \epsilon_\beta g_{\alpha \gamma}+Z_3^1 \epsilon_\gamma g_{\alpha \beta}
\\&+Z_4^1 \epsilon_\alpha p_\beta p_\gamma+Z_5^1 \epsilon_\beta p_\alpha p_\gamma
\\&+Z_6^1 \epsilon_\gamma p_\alpha p_\beta  \,,\\
 \left\langle 0\left|J_{\alpha \beta \gamma}\right| 1^{(-P) C}(p)\right\rangle&=Z_7^1 \varepsilon_{\alpha \beta \gamma \tau} \epsilon^\tau+Z_8^1 \varepsilon_{\alpha \beta \tau \lambda} \epsilon^\tau p^\lambda p_\gamma
 \\&+Z_9^1 \varepsilon_{\alpha \gamma \tau \lambda} \epsilon^\tau p^\lambda p_\beta  \,,\\
\left\langle 0\left|J_{\alpha \beta \gamma}\right| 2^{(-P) C}(p)\right\rangle&=Z_1^2 \epsilon_{\alpha \beta} p_\gamma+Z_2^2 \epsilon_{\alpha \gamma} p_\beta+Z_3^2 \epsilon_{\beta \gamma} p_\alpha  \,,\\
\left\langle 0\left|J_{\alpha \beta \gamma}\right| 2^{PC}(p)\right\rangle&=Z_4^2 \varepsilon_{\alpha \beta \tau \theta} \epsilon_\gamma^{~\tau} p^\theta+Z_5^2 \varepsilon_{\alpha \gamma \tau \theta} \epsilon_\beta^{~\tau} p^\theta  \,,\\
\left\langle 0\left|J_{\alpha \beta \gamma}\right| 3^{PC}(p)\right\rangle&=Z_1^3 \epsilon_{\alpha \beta \gamma} \, ,
\end{aligned}
\label{Eq:couplings}
\end{equation}
where $\epsilon_\alpha, \epsilon_{\alpha \beta}, \epsilon_{\alpha \beta \gamma}$ are the spin-1, spin-2 and spin-3 polarization tensors respectively. The parity in Eq.~\eqref{Eq:couplings} is $P=+$ for $J_{\alpha \beta \gamma }^{(1)}$ and $J_{\alpha \beta \gamma }^{(4)}$ while $P=-$ for $J_{\alpha \beta \gamma }^{(2)}$ and $J_{\alpha \beta \gamma }^{(3)}$.

For the interpolating currents in Eq.~\eqref{Eq:current}, they can not couple to a spin-3 state since the last two Lorentz indices in $G_{\beta\gamma}$ and $\tilde G_{\beta\gamma}$ are antisymmetric while the spin-3 polarization tensor $\epsilon_{\alpha \beta \gamma}$ is completely symmetric. According to Eq.~\eqref{Eq:couplings}, all these currents can couple to both positive and negative parity states with spin-0, spin-1 and spin-2 via different tensor structures. It is explicit that the charge conjugation $C$-parities of $J_{\alpha \beta \gamma }^{(1)}$ and $J_{\alpha \beta \gamma }^{(2)}$ are negative while those of $J_{\alpha \beta \gamma }^{(3)}$ and $J_{\alpha \beta \gamma }^{(4)}$ are positive. 

The couplings between a three Lorentz indices current and hadron states are very complicate, as shown in Eq.~\eqref{Eq:couplings}. For the currents $J_{\alpha \beta \gamma }^{(1)}$ and $J_{\alpha \beta \gamma }^{(2)}$, we concern the following couplings to study the hybrid states with $J^{PC}=2^{+-}$
\begin{equation}
\begin{aligned}
\left\langle 0\left|J_{\alpha \beta \gamma}^{(1)}\right| 2^{+-}(p)\right\rangle
=&Z_4^2 \varepsilon_{\alpha \beta \tau \theta} \epsilon_\gamma^{~\tau} p^\theta+Z_5^2 \varepsilon_{\alpha \gamma \tau \theta} \epsilon_\beta^{~\tau} p^\theta \\
=&f^{-}\left(\varepsilon_{\alpha \beta \tau \theta} \epsilon_\gamma^{~\tau} p^\theta-\varepsilon_{\alpha \gamma \tau \theta} \epsilon_\beta^{~\tau} p^\theta\right)\, ,
\end{aligned}
\label{Eq:coupling1}
\end{equation}
\begin{equation}
\begin{aligned}
\left\langle 0\left|J_{\alpha \beta \gamma}^{(2)}\right| 2^{+-}(p)\right\rangle =&Z_1^2 \epsilon_{\alpha \beta} p_\gamma+Z_2^2 \epsilon_{\alpha \gamma} p_\beta+Z_3^2 \epsilon_{\beta \gamma} p_\alpha \\
=&
f^{-\prime}(\epsilon_{\alpha \beta } p_\gamma  -\epsilon_{\alpha \gamma} p_\beta)\, ,
\end{aligned}
\label{Eq:coupling2}
\end{equation}
in which $f^-$ and $f^{-\prime}$ are the coupling constants between the currents and the corresponding $2^{+-}$ hybrids. In accord with the currents in Eq.~\eqref{Eq:current}, the tensor structures of the above relations are rewritten as antisymmetric for the indices $\{\beta\gamma\}$, so that the symmetric part vanishes and only one $2^{+-}$ state can be extracted from each interpolating current. Due to these couplings, one can prove that $J_{\alpha \beta \gamma }^{(1)}$ and $J_{\alpha \beta \gamma }^{(2)}$ actually couple to the same hybrid state, by multiplying a Levi-Civita tensor on both side of Eq.~\eqref{Eq:coupling1}, and thus the Lorentz structures for $f^-$ and $f^{-\prime}$ are dual conjugation of each other.
To study the hybrid mesons with $J^{PC}=2^{+-}$, we shall investigate the interpolating current $J_{\alpha \beta \gamma }^{(1)}$ in this work.

\section{QCD sum rules}
The two-point correlation function of $J_{\alpha \beta \gamma }^{(1)}$ can be written as
\begin{equation}
\begin{aligned}
 \Pi_{\alpha_{1}\beta_{1} \gamma_{1},\alpha_{2}\beta_{2} \gamma_{2}}(p^{2}) &=i \int d^{4} x e^{i p \cdot x}\left\langle 0\left|T\left[J_{\alpha_{1}\beta_{1} \gamma_{1}}^{(1)}(x) J_{\alpha_{2}\beta_{2} \gamma_{2}}^{(1)\dagger}(0)\right]\right| 0\right\rangle \, ,
 \label{Eq:correlator}
\end{aligned}
\end{equation}
which contains six Lorentz indices. To study the hybrid state with $J^{PC}=2^{+-}$, we decompose this correlation function into invariant function associated with such quantum numbers by constructing the normalized projector
\begin{equation}
\begin{aligned}
\mathbb{P}_{\alpha_{1},\beta_{1},\gamma_{1},\alpha_{2},\beta_{2},\gamma_{2}}=&\frac{1}{80p^2}\times\\
&\sum \left(\varepsilon_{\alpha_{1} \beta_{1}  \tau_{1}  \theta_{1} } \epsilon_{\gamma_{1} }^{~\tau_{1} } p^{\theta _{1}}-\varepsilon_{\alpha_{1}  \gamma_{1}  \tau_{1}  \theta_{1} } \epsilon_{\beta_{1} }^{~\tau_{1} } p^{\theta_{1} }\right)\\
&\times\left(\varepsilon_{\alpha_{2}  \beta_{2} \tau_{2} \theta_{2}} \epsilon_{\gamma_{2}}^{~\tau_{2}*} p^{\theta_{2}}-\varepsilon_{\alpha_{2} \gamma_{2} \tau_{2} \theta_{2}} \epsilon_{\beta_{2}}^{~\tau_{2}*} p^{\theta_{2}}\right)  \, .
\end{aligned}
\label{Eq:Projector1}
\end{equation}
The above summation over the polarization tensor is
\begin{equation}
\sum \epsilon_{\alpha_{1} \beta_{1}} \epsilon_{\alpha_{2} \beta_{2}}^{*}=\eta_{\alpha_{1} \alpha_{2}} \eta_{\beta_{1} \beta_{2}}+\eta_{\alpha_{1} \beta_{2}} \eta_{\beta_{1} \alpha_{2}}-\frac{2}{3} \eta_{\alpha_{1} \beta_{1}} \eta_{\alpha_{2} \beta_{2}} \,,
\end{equation}
where
\begin{equation}
\eta_{\alpha\beta}=\frac{p_{\alpha} p_{\beta}}{p^{2}}-g_{\alpha \beta} \,.
\end{equation}
One may wonder whether there is the Lorentz structure $\varepsilon_{\beta \gamma \tau \theta} \epsilon_\alpha^{~\tau} p^\theta $ in Eq.~\eqref{Eq:coupling1}. As a matter of fact, the projector constructed by this structure is the same as that in Eq.~\eqref{Eq:Projector1}. In other words, there are only two independent Lorentz structures in Eq.~\eqref{Eq:coupling1}. 

The invariant function $\Pi_{2}\left(p^{2}\right)$ corresponding to the $J^{PC}=2^{+-}$ can be extracted as
\begin{equation}
  \begin{aligned}
    \Pi_{2}(p^{2}) &=\mathbb{P}^{\alpha_{1},\beta_{1},\gamma_{1},\alpha_{2},\beta_{2},\gamma_{2}}\Pi_{\alpha_{1}\beta_{1} \gamma_{1},\alpha_{2}\beta_{2} \gamma_{2}}(p^{2}) \, .
   \label{Eq:Pi_2}
  \end{aligned}
  \end{equation}
This correlation function is usually evaluated via the method of operator production expansion (OPE) at the quark-gluon level, as a function of various QCD parameters and nonperturbative condensates. 

At the hadron level, the correlation function $\Pi(p^{2})$ can be described via the dispersion relation
\begin{equation}
\Pi(p^{2})=\frac{\left(p^{2}\right)^{N}}{\pi} \int_{0}^{\infty} \frac{\operatorname{Im} \Pi(s)}{s^{N}\left(s-p^{2}-i \epsilon\right)} d s+\sum_{n=0}^{N-1} b_{n}\left(p^{2}\right)^{n}\, ,
\label{Cor-Spe}
\end{equation}
where $b_n$ is the subtraction constant. In QCD sum rules, the imaginary part of the correlation function is defined as the spectral function
\begin{equation}
\rho (s)=\frac{1}{\pi} \text{Im}\Pi(s)=f_X^{2}\delta(s-m_{X}^{2})+\cdots\, ,
\end{equation}
in which the “one pole plus continuum” parametrization is used, and “$\cdots$" denotes the continuum and higher states. The parameters $f_X$ and $m_{X}$ are the coupling constant and mass of the lowest-lying hybrid meson respectively.

\begin{figure}[t!!]
  \centering
  \includegraphics[width=8cm,height=9cm]{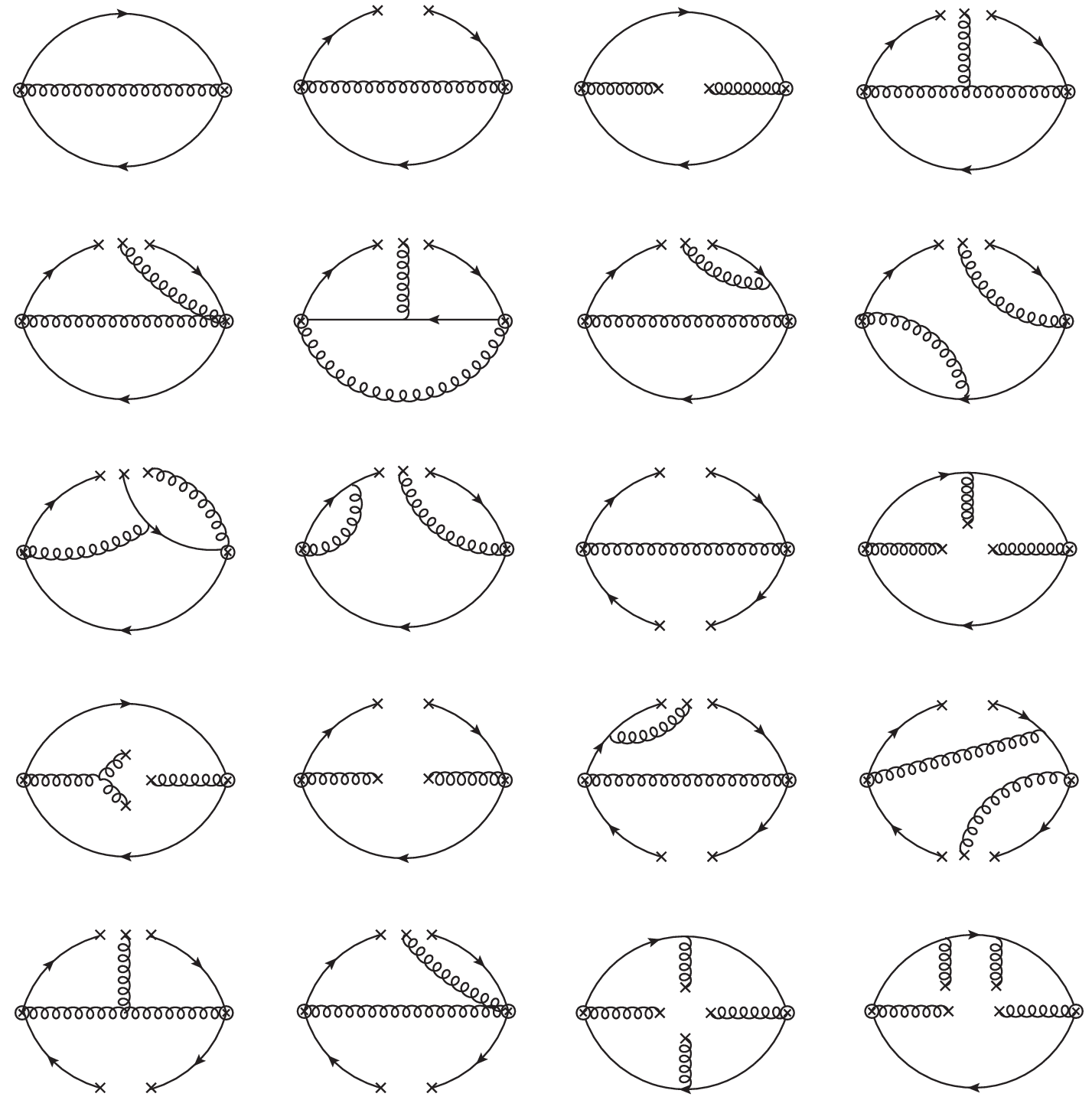}
\caption{Feynman diagrams evaluated for $2^{+-}$ hybrid states.}
  \label{fig: feyn_diagrams}
\end{figure}

To improve the convergence of the OPE series and suppress the contributions from continuum and higher states, the Borel transformation can be performed to the correlation function in both hadron and quark-gluon levels. The QCD sum rules are then obtained as
\begin{equation}
\Pi\left(s_{0}, M_{B}^{2}\right)=f_X^{2} e^{-m_{X}^{2} / M_{B}^{2}}=\int_{0}^{s_{0}} d s e^{-s / M_{B}^{2}} \rho(s)\, ,
\end{equation}
where $M_B$ is the Borel mass introduced via the Borel transformation and $s_0$ is the continuum threshold. 
Then the mass and coupling of the lowest-lying hadron can be extracted as
\begin{equation}
   \begin{aligned}
   m_{X}\left(s_0, M_B^{2}\right) & =\sqrt{\frac{\frac{\partial}{\partial\left(-1 / M_B^2\right)} \Pi\left(s_0, M_B^2\right)}{\Pi\left(s_0, M_B^2\right)}} \,,\\
  f_{X}\left(s_0, M_B^{2}\right) & =\sqrt{\Pi\left(s_{0}, M_{B}^{2}\right)e^{m_{X}^{2} / M_{B}^{2}}}\,. \label{mass+coupling}
 \end{aligned}
\end{equation}
In this work, we calculate the correlation function and spectral density at the leading order of $\alpha_{s}$ up to dimension eight condensates. The corresponding Feynman diagrams  are listed in Fig.~\ref{fig: feyn_diagrams}.
For the current $J_{\alpha \beta \gamma }^{(1)}$, the correlation function for the hybrid meson with $J^{PC}=2^{+-}$ is obtained as
\begin{eqnarray}
\nonumber    
\Pi_{2}(s_{0},M_{B}^{2})&=& \int_{0}^{s_{0}}ds~e^{-\frac{s}{M_{B}^{2}}}
    \left\{\frac{\alpha_{s}s^{3}}{2^{8}\times 3^{2}~ \pi^{3}} +\frac{11 \alpha_{s} m_{q} s }{2^{4}\times 3^{2}~ \pi}\dqq \right. \\ 
    \nonumber &&\left. +\frac{s }{2^{4}\times 3^{2}~ \pi}\dGG +\frac{139 \alpha_{s} m_{q}}{2^{7}\times 3^{3}~\pi}\dqGq \right. \\  \nonumber &&\left. +\frac{1}{2^{8}~\pi^{2}}\dGGG \right\}+\frac{19\pi \alpha_{s}}{2^{4} \times 3^{2} }\dqq \dqGq\\ \nonumber
  &&  -\frac{1}{2^{7}\times 3^{2} }\dGG^{2}+  \frac{\pi m_{q}}{18 }\dqq\dGG\\&&-\frac{5}{2^{8}\times 3^{4}}(\mathrm{ln}(M_{B}^{2}/\mu^{2})-\gamma_{E})\dGG^{2}\, .
\label{Eq:Result}
\end{eqnarray}
At the leading order of $\alpha_{s}$, the four-quark condensate gives no contribution to the correlation function since the corresponding feynman diagram has no loop~\cite{Ho:2018cat}. For the light $\bar{u}gu$ and $\bar{d} gd$ hybrid systems, we don't consider the contributions from the terms proportional to $m_q$ due to the negligible light quark masses. For the strangeonium $\bar{s}gs$ hybrid system, we take into account these $m_s$ terms up to dimension-7.
Generally speaking, the calculations for the dimension-7 and dimension-8 condensates are very complicated. One can consult the Refs.~\cite{Pimikov:2022brd,Grozin:1994hd} for the methods of such calculations. Considering the ignorable contributions of these terms, we just simply calculate the condensates $\dqq\dGG$, $\dqq\dqGq$ and $\dGG^2$ by applying the factorization assumption, which is usually adopted in QCD sum rules to estimate the values of the high dimension condensates~\cite{Shifman:1978bx,Reinders:1984sr}. 

\section{Numerical analysis}
To perform the numerical analyses, we use the following values for various QCD parameters at the renormalization scale $\mu=2 ~\mathrm{GeV}$ and the QCD scale $\Lambda_{QCD}=300 ~\mathrm{MeV}$~\cite{ParticleDataGroup:2022pth,Narison:2011xe,Narison:2018dcr,Jamin:2002ev}:
\begin{align}
\alpha_s \left(\mu\right) & =\frac{4 \pi}{9 \ln \left(\mu^{2} / \Lambda_{\mathrm{QCD}}^2\right)}\, , \nonumber \\
m_{s} & =93_{-5}^{+11}  ~\mathrm{MeV}\, , \nonumber\\
\dqq & =-(0.24 \pm 0.01)^3 ~\mathrm{GeV}^3 \,,  \nonumber\\
\dss & =(0.8\pm 0.1)\times \dqq \,,  \nonumber\\
\dqGq & =-(0.8 \pm 0.2) \times \dqq ~\mathrm{GeV}^2 \,,  \nonumber\\
\dsGs & =(0.8\pm 0.2) \times \dqGq\, ,\\ 
\dGG & =(6.35 \pm 0.35) \times 10^{-2} ~\mathrm{GeV}^4 \,,  \nonumber\\
\dGGG & =-(8.2 \pm 1.0) \times \dGG ~\mathrm{GeV}^2 \, . \nonumber
\end{align}
For the light quarks, we use the so-called ``current quark masses'' in a mass-independent subtraction scheme at a scale 2 GeV. The chiral limit is adopted so that $m_u=m_d=0$.  

As shown in Eq.~\eqref{mass+coupling}, the hadron mass and coupling are the function of $M_B$ and $s_0$. The working regions of these two parameters can be determined by requiring suitable OPE convergence and pole contribution.
To guarantee the good OPE convergence, we require that the contributions from the $D>6$ condensates be less than 2\%, i.e
\begin{equation}
R_{D>6}=\left|\frac{\Pi_2^{D>6}\left(M_B, \infty\right)}{\Pi_2^{t o t}\left(M_B, \infty\right)}\right|<2 \% \,.
\end{equation}
\begin{figure}[h!!]
  \centering
  \includegraphics[width=7cm]{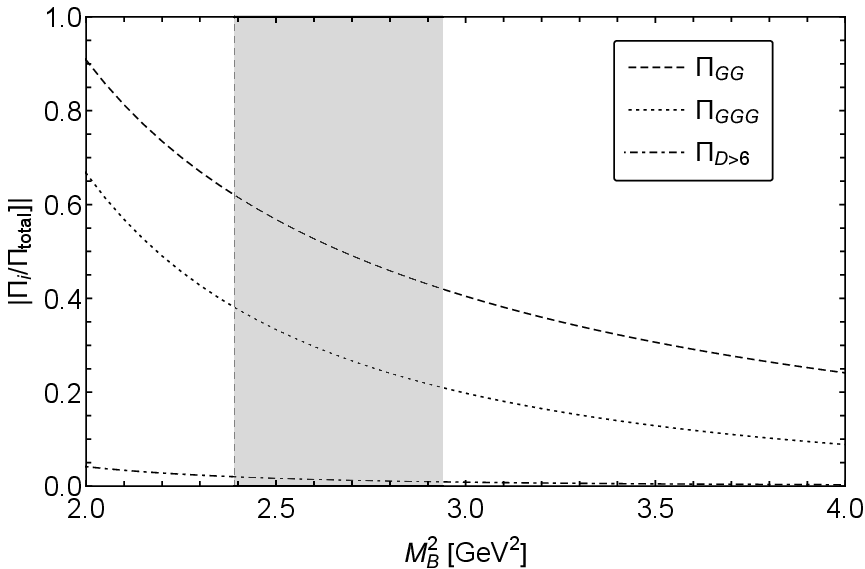}\\
 \caption{OPE convergence for the correlation function for the $J^{PC}=2^{+-}$ $\bar{q}gq$ hybrid state.}
 \label{fig:Convergence}
\end{figure}

\begin{figure}[h!!]
  \centering
  \subfigure[]{\includegraphics[width=4.2cm,height=3.2cm]{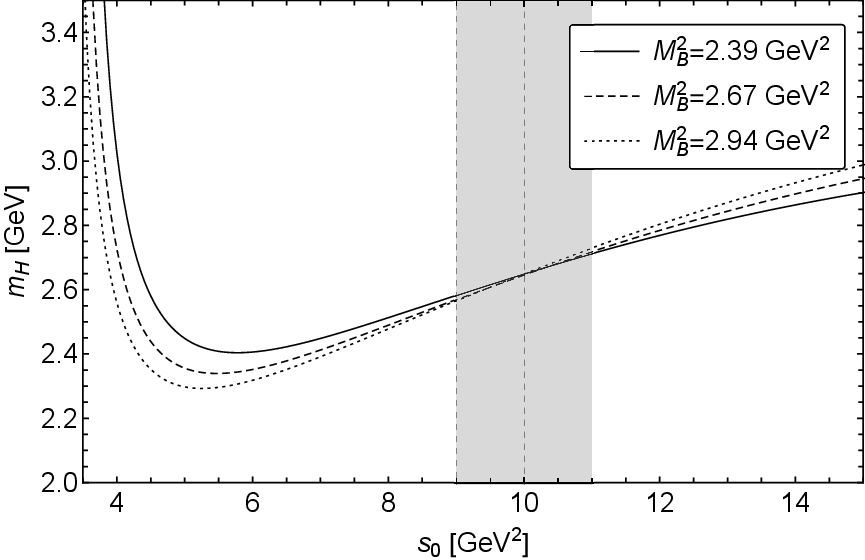}\label{fig:s0-mH}}
  \subfigure[]{\includegraphics[width=4.2cm,height=3.2cm]{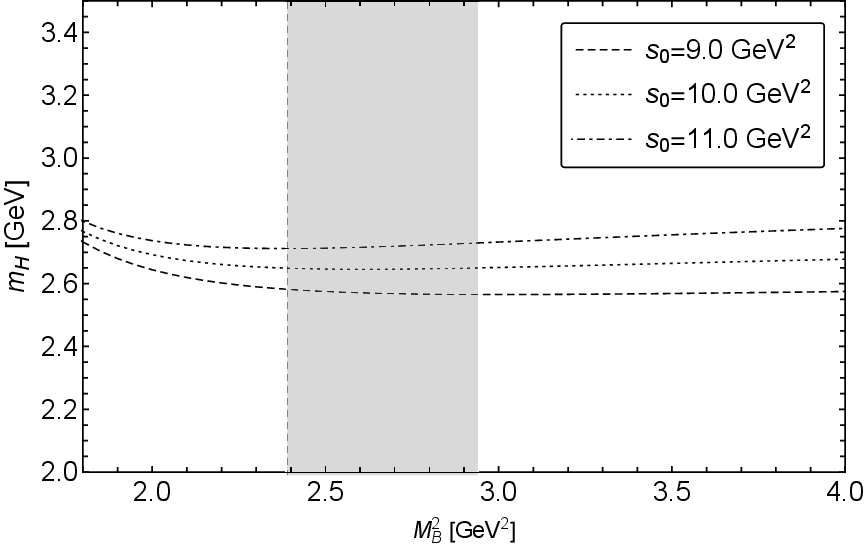}\label{fig:MB-mH}}\\
  \caption{Variations of hadron mass with $s_{0}$ and $M_{B}^{2}$ for the $\bar{q}gq$ light hybrid meson.}
  \label{fig:Result}
\end{figure}

For the $\bar qgq$ light hybrid meson with $J^{PC}=2^{+-}$, this requirement leads to the lower bound of the Borel mass $M_{B}^{2}\geq 2.39~\mathrm{GeV}^{2}$. We show the OPE convergence of the correlation function in Fig.~\ref{fig:Convergence}.
To get the upper bound of $M_{B}^{2}$, we need to fix the value of $s_{0}$ at first. In Fig.~\ref{fig:s0-mH}, we show the variations of the hadron mass $m_{X}$ with the threshold $s_{0}$ for various Borel mass $M_{B}^{2}$. It is shown that the variation of $m_{X}$ with  $M_{B}^{2}$ is minimized around $s_{0}\sim 10.0~\mathrm{GeV}^{2}$, which will result in the working region $9.0\leq s_0\leq 11.0~\mathrm{GeV}^{2}$. Using this value of $s_{0}$, the upper bound of $M_{B}^{2}$ can be obtained by requiring the pole contribution be larger than 50\%, i.e
\begin{equation}
 P C=\frac{\Pi\left(M_B, s_0\right)}{\Pi\left(M_B, \infty\right)}>50 \% \, .
\end{equation}

Finally, the working region of the Borel parameter can be determined to be $2.39\leq M_{B}^{2}\leq 2.94~\mathrm{GeV}^{2}$. We show the Borel curves in these regions in Fig.~\ref{fig:MB-mH}, from which we can see that the sum rule is stable in the above parameter regions. We can extract the hadron mass as
\begin{equation}
  \begin{aligned}
    m_{b_2/h_2}&=2.65\pm 0.09~\mathrm{GeV}\, , \label{qgqmass}
  \end{aligned}
\end{equation}
and the coupling as
\begin{equation}
  \begin{aligned}
    f_{b_2/h_2}&=(1.25\pm 0.18)\times 10^{-1}~\mathrm{GeV}^{4}\, .
  \end{aligned}
\end{equation}
The errors come from the continuum threshold $s_{0}$, condensates $\dGG$, $\dqGq$ and $\dGGG$. The uncertainty from the Borel mass is small enough to be neglected. The hadron mass extracted in Eq.~\eqref{qgqmass} is degenerate for the isoscalar and isovector $\bar qgq$ light hybrid states with $J^{PC}=2^{+-}$, since we don't distinguish them in our calculations. 

\begin{figure}[t!!]
  \centering
  \includegraphics[width=7cm]{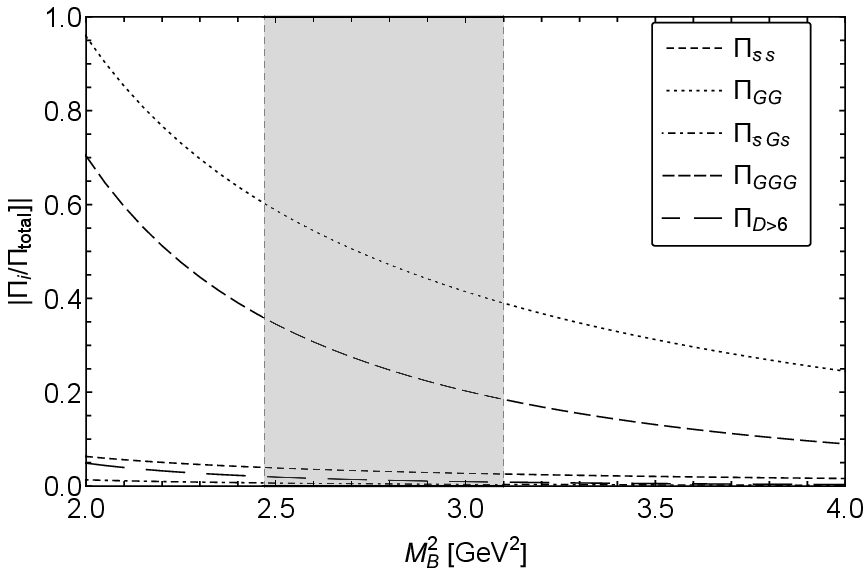}\\
 \caption{OPE convergence for the correlation function for the $J^{PC}=2^{+-}$ $\bar{s}gs$ hybrid state.}
 \label{fig:Convergence-s}
\end{figure}

\begin{figure}[h!!]
  \centering
  \subfigure[]{\includegraphics[width=4.2cm,height=3.2cm]{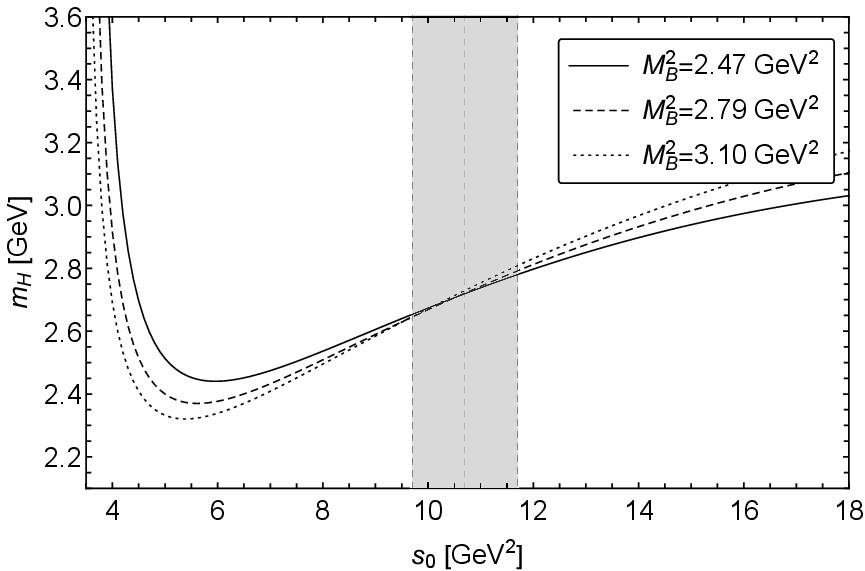}\label{fig:s0-mH-s}}
  \subfigure[]{\includegraphics[width=4.2cm,height=3.2cm]{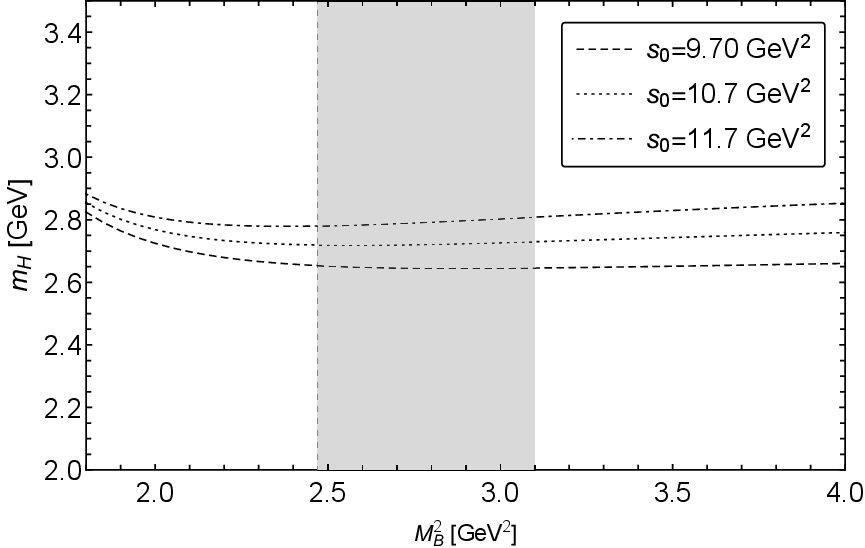}\label{fig:MB-mH-s}}\\
  \caption{Variations of hadron mass with $s_{0}$ and $M_{B}^{2}$ for the $\bar{s}gs$ strangeonium hybrid meson.}
  \label{fig:Result-s}
\end{figure}

For the strangeonium $\bar{s}g s$ hybrid system, the dimension-odd condensates proportional to $m_{s}$ are taken into account for the correlation function. However, their contributions are much smaller than those from the two- and tri-gluon condensates, as shown in Fig.~\ref{fig:Convergence-s}. Adopting the same criteria as the above mentioned, the Borel parameter and continuum threshold can be determined to be $2.47\leq M_{B}^{2}\leq 3.10~\mathrm{GeV}^{2}$ and $9.7\leq s_0\leq 11.7~\mathrm{GeV}^{2}$. We show the Borel curves in Fig.~\ref{fig:Result-s}.
The mass and coupling of this state are
\begin{equation}
  \begin{aligned}
    m_{h_2^\prime}&=2.72\pm 0.09~\mathrm{GeV}\, , \label{sgsmass}
  \end{aligned}
\end{equation}
and
\begin{equation}
  \begin{aligned}
    f_{h_2^\prime}&=(1.38\pm 0.18)\times 10^{-1}~\mathrm{GeV}^{4}\, .
  \end{aligned}
\end{equation}

Our calculations of the $2^{+-}$ hybrid masses in Eqs.~\eqref{qgqmass} and~\eqref{sgsmass} are in good agreement with the LQCD predictions~\cite{Dudek2011,Dudek2013}.

\section{Production and decay}
\begin{figure}[t!!]
  \centering
  \subfigure[]{\includegraphics[width=4.2cm,height=2.5cm]{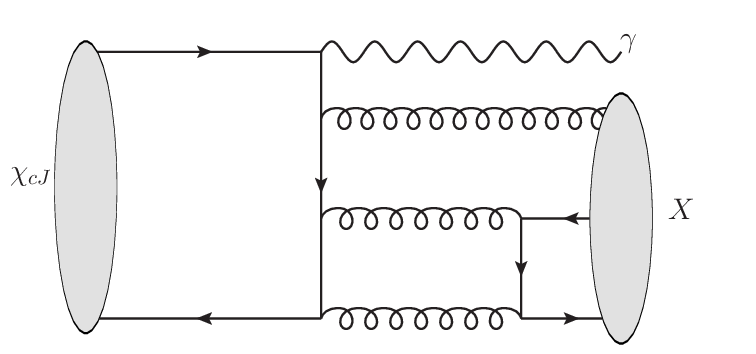}\label{fig:chi_decay1}}
  \subfigure[]{\includegraphics[width=4.2cm,height=2.5cm]{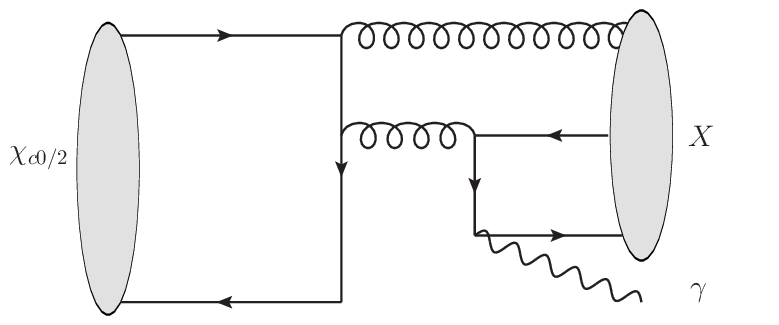}\label{fig:chi_decay2}}
  \caption{Two possible production mechanisms of $2^{+-}$ light hybrids from $\chi_{cJ}$ radiative decays.}
  \label{fig: production}
\end{figure}

Similar to the $1^{-+}$ hybrid candidate $\eta_{1}(1855)$ being observed in the radiative decays of $J/\psi$\cite{BESIII:2022riz,Chen:2022isv}, a light hybrid meson $X$ with $J^{PC}=2^{+-}$ can be generated from the decay of charmoniums, such as  $\chi_{cJ} \rightarrow \gamma X$. For this reason, $\psi(3686)\rightarrow \gamma \chi_{cJ}\rightarrow \gamma  \gamma X$ is a possible process to produce $2^{+-}$ light hybrid mesons. Analogous to the radiative decay $J/\psi\rightarrow \gamma\eta_{1}(1855)$, the decay $\chi_{cJ} \rightarrow \gamma X$ can also occur through the three-gluon emission processes as shown in Fig.~\ref{fig:chi_decay1}. In additon, the $2^{+-}$ light hybrids can be generated through the two-gluon emission process either, as shown in Fig.~\ref{fig:chi_decay2}, since the $\chi_{c0/2}$ can decay in such a process. For this reason, the production rates of the $2^{+-}$ hybrids in $\chi_{cJ}$ radiative decays may be larger than that of $\eta_{1}(1855)$ in $J/\psi$ radiative decays. This advantage can be utilized to search for these hybrid mesons from the high-statistics samples of $\psi(3686)$ in BESIII and BelleII experiments.

A hybrid meson may decay into two conventional mesons in two ways, one of which is that a pair of $\bar{q}q$ or $\bar{s}s$ is excited from the valence gluon and then combines with the valence quark and antiquark respectively to form two mesons. Another way is the so-called QCD axial anomaly~\cite{Akhoury:1987ed,Ball:1995zv,Chao:1989yp}.
Both mechanisms are at the order of $\alpha_{s}$, and thus may be on the same order of magnitude. Meanwhile, a hybrid meson can also decay into final states containing a lighter hybrid meson plus a conventional meson, for example, the S-wave decay processes $b_2\to\pi_1\omega/\eta_1\rho$, $h_2\to\eta_1\omega/\pi_1\rho$. 

We list some possible two-meson decay modes of $2^{+-}$ hybrids with different $I^{G}$ in Table~\ref{tab:decay_mode}, in which the S-wave, P-wave and D-wave decay channels are considered. It is clear that all S-wave decay channels have limited phase spaces, while P-wave channels such as  $a_{1}\pi$, $b_{1}\pi$, $h_{1}\pi$, $a_{2}\pi$, $K\bar{K}_{1}$ have larger phase spaces. It is interesting to find that all two S-wave mesons decay channels 
are in D-wave, such as $\omega\pi$, $\rho\eta^{(\prime)}$, $\rho\pi$, $\omega\eta^{(\prime)}$, $K\bar{K}^\ast$, with much larger phase spaces than those of S-wave and P-wave channels. Such decay behaviors may result in the small total decay width for these $2^{+-}$ hybrid mesons~\cite{Page:1998gz}, making it possible to isolate and detect them. Accordingly, we suggest to search for the $b_2$ hybrid in the $4\pi$ final states via $b_2\to\omega/a_1/h_1/a_2\pi\to4\pi$ process, while the $h_2$ hybrid in the $3\pi$ or $5\pi$ final states via $h_2\to\rho\pi\to3\pi$ or $h_2\to b_1\pi\to5\pi$ process. The two strange mesons channels may also be used to reproduce these hybrid states, such as $K\bar{K}^\ast, K^{(\ast)}\bar K_1$ and so on. 

Besides, these $2^{+-}$ hybrid mesons can also decay into the $\pi \gamma$ and $\eta^{(\prime)} \gamma$ final states via the electromagnetic interaction. 

\begin{table}[t!]
  \caption{Some possible two-meson decay modes for the hybrids with the exotic quantum numbers $I^GJ^{PC}=0^-2^{+-}$ ($h_2$ and $h_2^\prime$) and $1^+2^{+-}$ ($b_2$). The two strange mesons decay modes are appropriate for $b_2, h_2$ and $h_2^\prime$ at the same time.}
  \renewcommand\arraystretch{1.3} 
  \setlength{\tabcolsep}{1.5em}{ 
  \begin{tabular}{ccc}
   \hline  \hline 
    $I^{G}(J^{PC})$     & $0^{-}(2^{+-})$    &$1^{+}(2^{+-})$ \\ \hline
    \multirow{2}{*}{S-wave }  & \multicolumn{2}{c}{$K_{0}^{*}\bar{K}_{2}^{*}$} \\  
                              &$a_{1}b_{1}$,\,$f_{1}h_{1}$,\,$f_{2}h_{1}$,& $a_{1}h_{1}$,\,$f_{1}b_{1}$,\,$f_{2}b_{1}$,\\&$\eta_1\omega$,\,$\pi_1\rho$ &$\pi_1\omega$,\,$\eta_1\rho$\\ \cline{2-3}
   \multirow{3}{*}{P-wave}   & \multicolumn{2}{c}{$K\bar{K}_{1}$,\,$K\bar{K}_{2}^{*}$,\,$K^{*}\bar{K}_{0}^{*}$,\,$K^{*}\bar{K}_{1}$,\,$K^{*}\bar{K}_{2}^{*}$}   \\  
   & $h_{1}\eta$,\,$b_{1}\pi$,\,$f_{0}\omega$\,,& $b_{1}\eta$,\,$h_{1}\pi$,\,$f_{0}\rho$\,,\\ & $f_{1}\omega$,\,$f_{2}\omega$,\,$a_{0}\rho$,\,$a_{1}\rho$  & $f_{1}\rho$,\,$f_{2}\rho$,\,$a_{0}\omega$,\,$a_{1}\pi$,\,$a_{2}\pi$\\
   \cline{2-3}
   \multirow{3}{*}{D-wave}   & \multicolumn{2}{c}{$K\bar{K}^{\ast}$}   \\  
   & $\rho\pi$,\,$\omega\eta^{(\prime)}$& $\omega\pi$,\,$\rho\eta^{(\prime)}$\\
       \hline\hline  
  \label{tab:decay_mode}
  \end{tabular}}
\end{table}

\section{Conclusion}
We have investigated the masses of the light hybrid mesons with exotic quantum number $J^{PC}=2^{+-}$ in QCD sum rule method by constructing local  interpolating currents with three Lorentz indices. To calculate the correlation functions and spectral densities, we consider the contributions of nonperturbative effect up to dimension eight condensates at the leading order of $\alpha_{s}$. 

After the numerical analyses, the extracted masses are about $2.65~\mathrm{GeV}$ for the non-strange $\bar{q}gq$ hybrids ($b_2$ and $h_2$) and $2.72~\mathrm{GeV}$ for the $\bar{s}gs$ strangeonium hybrid ($h_2^{\prime}$), respectively. As discussed above, their peculiar decay behaviors may result in the surprising narrowness of these $2^{+-}$ hybrids. We suggest to hunt for such hybrid states through the partial wave analyses in the $b_2\to(\omega/a_1/h_1/a_2)\pi\to4\pi$, $h_2\to\rho\pi\to3\pi$ and $h_2\to b_1\pi\to5\pi$ processes in BESIII and BelleII experiments by using the high-statistics data samples of $\psi(3686)$.

\section*{ACKNOWLEDGMENTS}

This project is supported by the National Natural Science Foundation of China under Grants No. 12175318, the National Key Research and Development Program of China (2020YFA0406400), the Natural Science Foundation of Guangdong Province of China under Grant No. 2022A1515011922, the Fundamental Research Funds for the Central Universities.



\begin{thebibliography}{100}
  \bibitem{Chen:2016qju}
  H.~X.~Chen, W.~Chen, X.~Liu and S.~L.~Zhu,
  Phys. Rept. \textbf{639},1 (2016)
  
  \bibitem{Esposito:2016noz}
  A.~Esposito, A.~Pilloni and A.~D.~Polosa,
  Phys. Rept. \textbf{668},1 (2017)
  
  \bibitem{Guo:2017jvc}
  F.~K.~Guo, C.~Hanhart, U.~G.~Mei\ss{}ner, Q.~Wang, Q.~Zhao and B.~S.~Zou,
  Rev. Mod. Phys. \textbf{90}, 015004 (2018)
  
  \bibitem{Liu:2019zoy}
  Y.~R.~Liu, H.~X.~Chen, W.~Chen, X.~Liu and S.~L.~Zhu,
  Prog. Part. Nucl. Phys. \textbf{107},237 (2019)
  
  \bibitem{Brambilla:2019esw}
  N.~Brambilla, S.~Eidelman, C.~Hanhart, A.~Nefediev, C.~P.~Shen, C.~E.~Thomas, A.~Vairo and C.~Z.~Yuan,
  Phys. Rept. \textbf{873}, 1 (2020)
  
  \bibitem{Chen:2022asf}
  H.~X.~Chen, W.~Chen, X.~Liu, Y.~R.~Liu and S.~L.~Zhu,
  Rept. Prog. Phys. \textbf{86}, 026201 (2023)
  
  \bibitem{Meng:2022ozq}
  L.~Meng, B.~Wang, G.~J.~Wang and S.~L.~Zhu,
  Phys. Rept. \textbf{1019}, 1 (2023)
  
  \bibitem{IHEP-Brussels-LosAlamos-AnnecyLAPP:1988iqi}
  D.~Alde et~al.,
  Phys. Lett. B \textbf{205}, 397 (1988)
  
  \bibitem{E852:2001ikk}
  E.~I. Ivanov et~al.,
  Phys. Rev. Lett. \textbf{86}, 3977 (2001)
  
  \bibitem{E852:2004gpn}
  J.~Kuhn et~al.,
  Phys. Lett. B \textbf{595}, 109 (2004)
  
  \bibitem{JPAC:2018zyd}
  A.~Rodas et~al.,
  Phys. Rev. Lett. \textbf{122}, 042002 (2019)
  
  \bibitem{COMPASS:2014vkj}
  C.~Adolph et~al.,
  Phys. Lett. B \textbf{740}, 303 (2015), [Erratum: Phys. Lett. B  \textbf{811}, 135913 (2020)]
  
  \bibitem{BESIII:2022riz}
  M.~Ablikim et~al.,
  Phys. Rev. Lett. \textbf{129}, 192002 (2022)
  
  \bibitem{BESIII:2022iwi}
  M.~Ablikim et~al.,
  Phys. Rev. D \textbf{106}, 072012 (2022), [Erratum: Phys. Rev. D  \textbf{107}, 079901 (2023)]
  
  \bibitem{Chen:2022qpd}
  H.~X. Chen, N.~Su, and S.~L. Zhu,
  Chin. Phys. Lett \textbf{39}, 051201 (2022)
  
  \bibitem{Chen:2023ukh}
  B.~Chen, S.~Q. Luo, and X.~Liu,
  arXiv:2302.06785
  
  \bibitem{Chen:2022isv}
  F.~Chen et~al.,
  Phys. Rev. D \textbf{107}, 054511 (2023)
  
  \bibitem{Qiu:2022ktc}
  L.~Qiu and Q.~Zhao,
  Chin. Phys. C \textbf{46}, 051001 (2022)
  
  \bibitem{Shastry:2022upd}
  V.~Shastry,
  PoS \textbf{ICHEP2022}, 779 (2022)
  
  \bibitem{Shastry:2023ths}
  V.~Shastry and F.~Giacosa,
  arXiv:2302.07687
  
  \bibitem{Shastry:2022mhk}
  V.~Shastry, C.~S. Fischer, and F.~Giacosa,
  Phys. Lett. B \textbf{834}, 137478 (2022)
  
  \bibitem{Huang:2022tpq}
  Y.~Huang and H.~Q. Zhu,
  arXiv:2209.02879
  
  \bibitem{Yan:2023vbh}
  M.-J. Yan, J.~M. Dias, A.~Guevara, F.~K. Guo, and B.~S. Zou,
  Universe \textbf{9}, 109  (2023)
  
  \bibitem{Yang:2022rck}
  F.~Yang, H.~Q. Zhu, and Y.~Huang,
  Nucl. Phys. A \textbf{1030}, 122571 (2023)
  
  \bibitem{Barnes:1982tx}
  T.~Barnes, F.~E. Close, and F.~de~Viron,
  Nucl. Phys. B \textbf{224}, 241 (1983)
  
  \bibitem{Chanowitz:1982qj}
  M.~S. Chanowitz and S.~R. Sharpe,
  Nucl. Phys. B \textbf{222}, 211 (1983), [Erratum: Nucl. Phys. B \textbf{228}, 588 (1983)]
  
  \bibitem{HadronSpectrum:2012gic}
  L.~Liu et~al.,
  JHEP \textbf{07}, 126 (2012)
  
    \bibitem{Chen:2013zia}
  W.~Chen et~al.,
  JHEP \textbf{09}, 019 (2013)
  
  \bibitem{Meyer:2015eta}
  C.~Meyer and E.~Swanson,
  Prog. Part. Nucl. Phys. \textbf{82}, 21 (2015)
  
  \bibitem{Guo:2008yz}
  P.~Guo, A.~P. Szczepaniak, G.~Galata, A.~Vassallo, and E.~Santopinto,
  Phys. Rev. D \textbf{78}, 056003 (2008)

  \bibitem{Dudek2011}
  J.~J. Dudek,
  Phys. Rev. D \textbf{84}, 074023 (2011)
 
  \bibitem{Dudek:2009qf}
  J.~J. Dudek, R.~G. Edwards, M.~J. Peardon, D.~G. Richards and C.~E. Thomas,
  Phys. Rev. Lett \textbf{103}, 262001 (2009)
 
  \bibitem{Dudek:2011tt}
 J.~J.~Dudek, R.~G.~Edwards, B.~Joo, M.~J.~Peardon, D.~G.~Richards and C.~E.~Thomas,
  Phys. Rev. D \textbf{83}, 111502 (2011)

\bibitem{Dudek:2010wm}
  J.~J. Dudek, R.~G. Edwards, M.~J. Peardon, D.~G. Richards and C.~E. Thomas,
  Phys. Rev. D \textbf{82}, 034508 (2010)
  
  
  \bibitem{Lacock1997}
  P.~Lacock, C.~Michael, P.~Boyle, and P.~Rowland,
  Phys. Lett. B \textbf{401}, 308  (1997)
  
  \bibitem{Lacock1996}
  P.~Lacock, C.~Michael, P.~Boyle, and P.~Rowland,
  Phys. Rev. D \textbf{54}, 6997  (1996)
  
  \bibitem{Ma:2020bex}
  Y.~Ma, Y.~Chen, M.~Gong, and Z.~Liu,
  Chin. Phys. C \textbf{45}, 013112 (2021)
  

  \bibitem{Dudek:2009kk}
  J.~J. Dudek, R.~Edwards, and C.~E. Thomas,
  Phys. Rev. D \textbf{79}, 094504 (2009)
  
  \bibitem{Dudek2013}
  J.~J. Dudek, R.~G. Edwards, P.~Guo, and C.~E. Thomas,
  Phys. Rev. D \textbf{88},  094505 (2013)
  
  \bibitem{Isgur:1984bm}
  N.~Isgur and J.~E. Paton,
  Phys. Rev. D \textbf{31}, 2910 (1985)
  
  \bibitem{Isgur:1985vy}
  N.~Isgur, R.~Kokoski, and J.~Paton,
  Phys. Rev. Lett.  \textbf{54}, 869 (1985)
  
  \bibitem{Burns2006}
  T.~J. Burns and F.~E. Close
  Phys. Rev. D \textbf{74}, 034003 (2006)
  
  \bibitem{Burden2002}
  C.~J. Burden and M.~A. Pichowsky,
  Few Body Syst. \textbf{32}, 119 (2002)
  
  \bibitem{Burden:1996nh}
  C.~J. Burden, L.~Qian, C.~D. Roberts, P.~C. Tandy, and M.~J. Thomson, Phys. Rev. C \textbf{55}, 2649 (1997)
  
  \bibitem{Balitsky:1982ps}
  I.~I. Balitsky, D.~Diakonov, and A.~V. Yung,
  Phys. Lett. B \textbf{112}, 71 (1982)
  
  \bibitem{Latorre:1984kc}
  J.~I. Latorre, S.~Narison, P.~Pascual, and R.~Tarrach,
  Phys. Lett. B \textbf{147},  169 (1984)
  
  \bibitem{Govaerts:1983ka}
  J.~Govaerts, F.~de~Viron, D.~Gusbin, and J.~Weyers,
  Phys. Lett. B \textbf{128}, 262  (1983), [Erratum: Phys. Lett. B 136, 445 (1984)]
    
  \bibitem{Govaerts:1984bk}
  J.~Govaerts, F.~de~Viron, D.~Gusbin, and J.~Weyers,
  Nucl. Phys. B \textbf{248}, 1  (1984)
  
  \bibitem{Govaerts:1984hc}
  J.~Govaerts, L.~J. Reinders, H.~R. Rubinstein, and J.~Weyers,
  Nucl. Phys. B \textbf{258}, 215 (1985)

    \bibitem{Govaerts:1986pp}
  J.~Govaerts, L.~J. Reinders, P.~Francken, X.~Gonze, and J.~Weyers,
  Nucl. Phys. B \textbf{284}, 674 (1987)
  
  \bibitem{Balitsky:1986hf}
  I.~I. Balitsky, D.~Diakonov, and A.~V. Yung,
  Z. Phys. C \textbf{33}, 265 (1986)
  
  \bibitem{Barsbay:2022gtu}
  B.~Barsbay, K.~Azizi, and H.~Sundu,
  Eur. Phys. J. C \textbf{82}, 1086 (2022)
  
  \bibitem{Chen:2021smz}
  H.~X. Chen, W.~Chen, and S.~L. Zhu,
  Phys. Rev. D \textbf{105}, L051501 (2022)
  
    \bibitem{Chen:2013eha}
  W.~Chen, T.~G. Steele, and S.~L. Zhu,
  J. Phys. G \textbf{41}, 025003 (2014)
  
  \bibitem{Ho:2018cat}
  J.~Ho, R.~Berg, T.~G. Steele, W.~Chen, and D.~Harnett,  Phys. Rev. D \textbf{98},  096020 (2018)
  
  \bibitem{Ho:2019org}
  J.~Ho, R.~Berg, T.~G. Steele, W.~Chen, and D.~Harnett,
  Phys. Rev. D \textbf{100},  034012 (2019)
  
  \bibitem{Palameta:2018yce}
  A.~Palameta, D.~Harnett, and T.~G. Steele,
  Phys. Rev. D \textbf{98}, 074014 (2018)
  
  \bibitem{Page:1998gz}
  P.~R. Page, E.~S. Swanson, and A.~P. Szczepaniak,
  Phys. Rev. D \textbf{59}, 034016 (1999)
  
\bibitem{Shifman:1978bx}
M.~A.~Shifman, A.~I.~Vainshtein and V.~I.~Zakharov,
Nucl. Phys. B \textbf{147}, 385 (1979)

\bibitem{Reinders:1984sr}
L.~J.~Reinders, H.~Rubinstein and S.~Yazaki,
Phys. Rept. \textbf{127}, 1 (1985)

\bibitem{Colangelo:2000dp}
P.~Colangelo and A.~Khodjamirian,
arXiv:hep-ph/0010175

\bibitem{Nielsen:2009uh}
M.~Nielsen, F.~S.~Navarra and S.~H.~Lee,
Phys. Rept. \textbf{497}, 41 (2010)


\bibitem{Narison:2022paf}
S.~Narison,
Nucl. Part. Phys. Proc. \textbf{324-329}, 94 (2023)

\bibitem{Pimikov:2022brd}
A.~V.~Pimikov,
Phys. Rev. D \textbf{106}, 056011 (2022)

\bibitem{Albuquerque:2020hio}
R.~M.~Albuquerque, S.~Narison, A.~Rabemananjara, D.~Rabetiarivony and G.~Randriamanatrika,
Phys. Rev. D \textbf{102}, 094001 (2020)

\bibitem{Albuquerque:2021tqd}
R.~M.~Albuquerque, S.~Narison and D.~Rabetiarivony,
Phys. Rev. D \textbf{103}, 074015 (2021)

\bibitem{Wu:2022qwd}
R.~H.~Wu, Y.~S.~Zuo, C.~Y.~Wang, C.~Meng, Y.~Q.~Ma and K.~T.~Chao,
JHEP \textbf{11}, 023 (2022)
  
  \bibitem{Meyer:2010ku}
  C.~A. Meyer and Y.~Van~Haarlem,
  Phys. Rev. C \textbf{82}, 025208 (2010)


  
  \bibitem{ParticleDataGroup:2022pth}
  R.~L. Workman et~al.,
  PTEP \textbf{2022}, 083C01 (2022)

 \bibitem{Grozin:1994hd}
 A.~G.~Grozin,
 Int. J. Mod. Phys. A \textbf{10}, 3497 (1995)
  
  \bibitem{Narison:2011xe}
  S.~Narison,
  Phys. Lett. B \textbf{706}, 412 (2012)
  
  \bibitem{Narison:2018dcr}
  S.~Narison,
  Int. J. Mod. Phys. A \textbf{33}, 1850045 (2018)
  
  \bibitem{Jamin:2002ev}
  M.~Jamin,
  Phys. Lett. B \textbf{538}, 71 (2002)
  
  \bibitem{Akhoury:1987ed}
  R.~Akhoury and J.~M. Frere,
  Phys. Lett. B \textbf{220}, 258 (1989)
  
  \bibitem{Ball:1995zv}
  P.~Ball, J.~M. Frere, and M.~Tytgat,
  Phys. Lett. B \textbf{365}, 367 (1996)
  
  \bibitem{Chao:1989yp}
  K.~T. Chao,
  Nucl. Phys. B \textbf{317}, 597 (1989)
  
  \end{thebibliography}
\end{document}